\documentclass[prc,aps,twocolumn,showpacs,nofootinbib]{revtex4-1}%-1}
\usepackage{graphicx}% Include figure files
\newcommand{\be}{\begin{equation}}
\newcommand{\ee}{\end{equation}}
\newcommand{\ba}{\begin{array}}
\newcommand{\ea}{\end{array}}
\newcommand{\gnp}{\gamma n\!\to\!\pi^- p}

\newcommand{\gdpp}{\gamma d\!\to\!\pi^- pp}

\newcommand{\crs}{cross section}
\newcommand{\crss}{\crs s}

\def\pcomma{$^,$}
\def\pgwu{$^1$}
\def\pitep{$^2$}
\def\pinfn{$^3$}
\begin{document}

%%%%%%%%%%%%%%%%%%%%%%%%%%%%%%%%%%%%%%%%%%%%%%%%%%
\title{Evaluation of the  $\gnp$  differential \crs\ in the 
	$\Delta$-isobar region}

\author{W.~J.~Briscoe\pgwu,
	A.~E.~Kudryavtsev\pitep\pcomma\pgwu,
	P.~Pedroni\pinfn,
	I.~I.~Strakovsky\pgwu,
	V.~E.~Tarasov\pitep,
	R.~L.~Workman\pgwu\\ 
\vspace*{0.3in}}
%%%%%%%%%%%%%%%%%%%%%%%%%%%%%%%%%%%%%%%%%%%%%%%%%%%%%%%%%%%%%%%%%%
\affiliation{
\pgwu  The George Washington University, Washington, DC 20052, USA}
\affiliation{
\pitep Institute of Theoretical and Experimental Physics, Moscow, 
       117259 Russia}
\affiliation{
\pinfn INFN, Sezione di Pavia, via Bassi 6, 27100 Pavia , Italy\\}

\date{\today}

%%%%%%%%%%%%%%%%%%%%%%%%%%%%%%%%%%%%%%%%%%%%%%%%%%%%%%%%%%%%%%%%%%
\begin{abstract}
Differential cross sections for the process $\gnp$ have been 
extracted from MAMI-B measurements of $\gamma d\to \pi^- pp$, 
accounting for final-state interaction effects, using a 
diagrammatic technique taking into account the $NN$ and 
$\pi$N final-state interaction amplitudes. Results are 
compared to previous measurements of the inverse process, 
$\pi^- p\to n\gamma$, and recent multipole analyses.
\end{abstract}

\maketitle

%%%%%%%%%%%%%%%%%%%%%%%%%%%%%%%%%%%%%%%%%%%%%%%%%%%%%5
\section{Introduction}
\label{sec:intro}

An accurate evaluation of the electromagnetic (EM) couplings 
$N^\ast (\Delta^\ast)\to\gamma N$ from meson photoproduction 
data remains a paramount task in hadron physics. A wealth of 
new data for meson photoproduction is becoming available from
nuclear facilities worldwide. These measurements are now
beginning to have a significant impact on both the resonance
spectrum and its decay properties. 

Here we focus on the single-pion production data and note 
that a complete solution requires couplings from both charged 
and neutral resonances, the latter requiring  $\pi^-p$ and 
$\pi^0n$ photoproduction off a neutron target, typically a 
neutron bound in a deuteron target. Extraction of the two-body 
($\gamma n\to \pi^-p$ and $\gamma n\to\pi^0 n$) cross sections 
requires the use of a model-dependent nuclear correction, 
which mainly come from final state interactions (FSI). As a 
result, our knowledge of the neutral resonance couplings is 
less precise as compared to the charged values~\cite{PDG}.

In addition to being less precise, experimental data for
neutron-target photoreactions are much less abundant than 
those utilizing a proton target, constituting only about 15\% 
of the present SAID data base~\cite{SAID}. At low to 
intermediate energies, this lack of neutron-target data is 
partially compensated by experiments using pionic beams, e.g., 
$\pi^-p\to\gamma n$, as has been measured, for example, by 
the Crystal Ball Collaboration at BNL~\cite{aziz} for the 
inverse photon energy $E_\gamma$=285 -- 689~MeV and 
$\theta$=41 -- 148$^\circ$, where $\theta$ is the inverse 
production angle of $\pi^-$ in the center-of=mass (CM) frame.  
This process is free from complications associated with the 
deuteron target.  However, the disadvantage of using the 
reaction $\pi^-p\to\gamma n$ is the 5 to 500 times larger 
cross sections for $\pi^-p\to\pi^0n\to\gamma\gamma n$, 
depending on $E_\gamma$ and $\theta$.

We recently applied our FSI corrections~\cite{FSI} to CLAS
$\gdpp$ data~\cite{CLAS} to get elementary cross sections for 
$\gnp$~\cite{data}.  The FSI correction factor for the CLAS 
kinematics was found to be small, $\Delta\sigma/\sigma <$10\%. 
However, these new cross sections departed significantly from 
our predictions, at the higher energies, and greatly modified 
the fit result.
  
The present paper is addressed to differential cross section 
measurements for $\gamma n\to\pi^-p$ in the $\Delta(1232)$-isobar 
region. At energies dominated by the Delta resonance, the isospin 
$3/2$ multipoles are constrained by extensive studies performed 
using proton targets. The forward peaking structure is due 
largely to the Born contribution, which is well known. As a 
result, one would expect models to give predictions within a 
tight range, which is confirmed in Figs.~\ref{fig:g1} and 
\ref{fig:g5}. 

The paper is organized as follows. In Sec.~\ref{sec:DB}, 
we present the new data set and compare it with previous data 
from hadronic facilities. Sec.~\ref{sec:FSI} is devoted to 
the nuclear corrections. Here, we give comments on the bound 
neutron and discuss the effect of final state interaction 
(FSI) corrections. In Sec.~\ref{sec:res}, we correct the 
new data for FSI and compare with previous hadronic data and 
with predictions based on previous multipole analyses. The 
results of a fit are presented and considered along with the 
prospect of future polarized measurements.

%%%%%%%%%%%%%%%%%%%%%%%%%%%%%%%%%%%%%%%%%%%%%%%%
\section{Data SET}
\label{sec:DB}

In 2010, the GDH and A2 Collaborations published~\cite{MAMI} 
the first measurement of the unpolarized and the helicity 
dependent differential cross section for the $\gdpp$ reaction 
in the $\Delta$-resonance region. 

The events from this reaction were selected by requiring the 
presence of one charged pion and of one or two protons within 
the detector acceptance (momentum threshold for protons and 
charged pions $\sim$270~MeV/$c$ and $\sim$80~MeV/$c$, 
respectively; full azimuthal acceptance and polar laboratory 
angular acceptance between 21$^\circ$ and 159$^\circ$)

The obtained  results consist of $126$ experimental points 
covering a $E_\gamma$ range from 301 to 455~MeV and a pion 
polar angular emission range in the CM system between 
$\theta=58^\circ$ and  $\theta=141^\circ$.

During the data analysis phase, a kinematic calculation was 
performed to evaluate the momentum and the emission angle 
of the undetected proton. The calculated momentum distribution 
was found to be almost equal to the Fermi momentum distribution 
expected from the deuteron wave function (see Fig.~1 of 
Ref.~\cite{MAMI}).
%%%%%%%%%%%%%%%%%%%%%%%%%%%%%%%%%%%%%%%%%%%%%%%%%%%%%
\begin{figure*}[th]
\centerline{
\includegraphics[height=1\textwidth, angle=90]{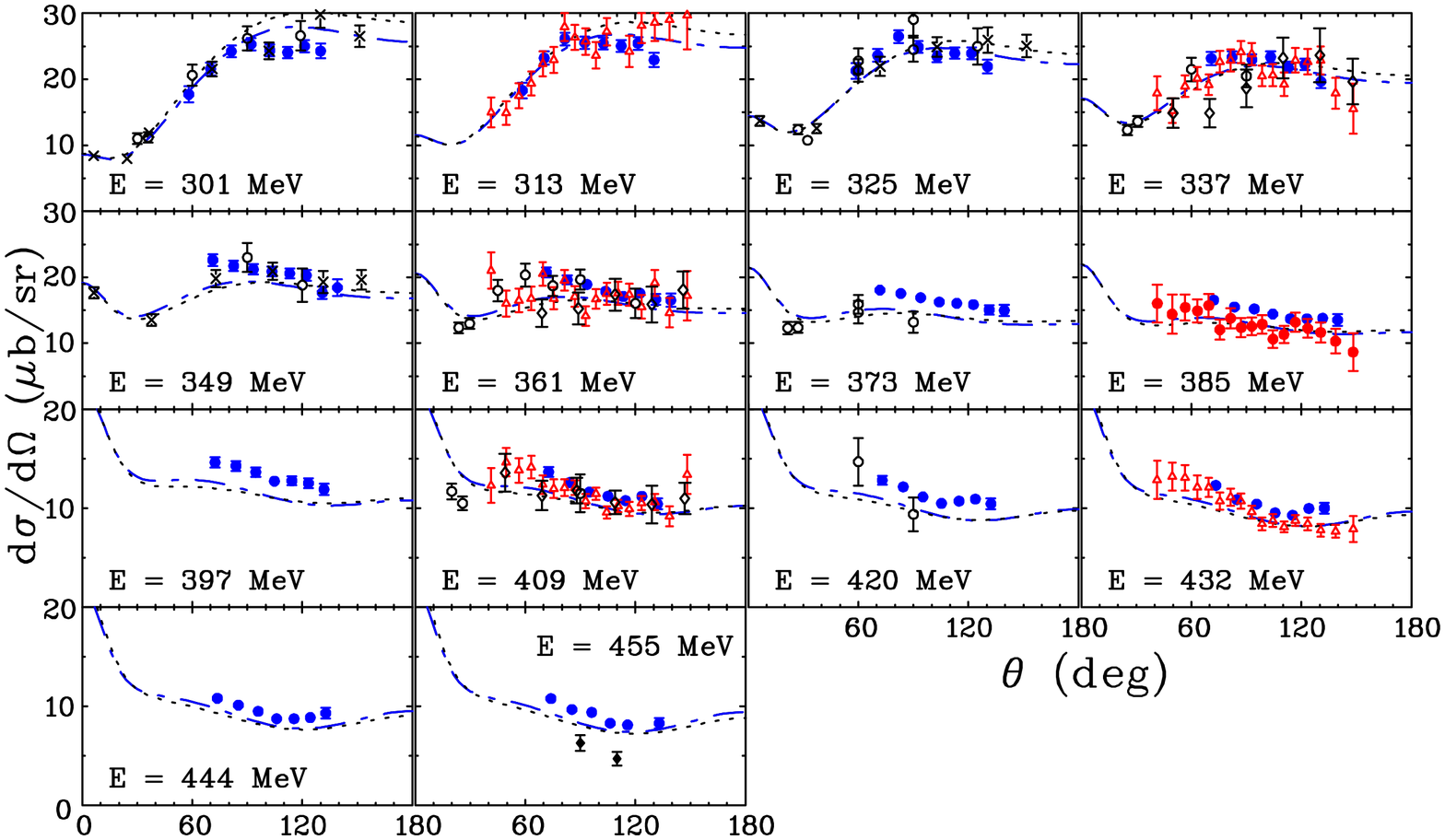}}
\caption{(Color online) Differential \crss\ for $\gnp$ as a
        function of $\theta$, the production angle of $\pi^-$ 
	in the CM frame. The present data (solid circles) are 
	shown for 14 energy bins.  Previous data came from
        MAMI-B~\protect\cite{MAMI} (blue filled circles),
        TRIUMF~\protect\cite{triumf} [296, 321, and 346~MeV]
                (black crosses)
        CERN~\protect\cite{cern} [301, 321, 339, 344, 356,
                376, 411, and 418] (black open circles),
        BNL~\protect\cite{aziz} [313, 338, 359, 390, 407, and
                436~~MeV] (red open triangles),
        LBL~\protect\cite{lbl} [335, 355, 360, and 409~MeV]
                (black open dimondes), and
        LAMPF~\protect\cite{lanl} [458~MeV] (black filled
        dimonds).
        Shown data came from hadronic facilities, except
        MAMI-B measurements (within $\Delta E_\gamma$ = 5~MeV
        binning).  Plotted uncertainties are statistical only.
        Blue dash-dotted (black dotted) lines correspond to
        the predictions for our recent SN11~\protect\cite{sn11}
        (MAID07~\protect\cite{MAID}) solution. \label{fig:g1}}
\end{figure*}
%%%%%%%%%%%%%%%%%%%%%%%%%%%%%%%%%%%%%%%%%%%%%%%%%%%%%%%

This comparison indicates that the dominant mechanism of 
the $\gdpp$ channel is the quasi-free reaction on the bound 
neutron while the proton acts merely as a spectator, remaining 
approximately at rest in the laboratory system.

MAMI-B deuteron data gives only the differential cross section
as function of the pion production laboratory angle and full
kinematics was not restored.  Thus, the $\gamma n\to\pi^- p$
cross section was extracted assuming the neutron to be at rest
to unambiguously relate the pion production angles $\theta$ in
the $\pi^- p$ CM rest frame to their measured laboratory angles.

Specific examples of agreement with previous measurements
are displayed in Fig.~\ref{fig:g1}, where we compare
differential \crss\ obtained here with those from hadronic
facilities (TRIUMF~\cite{triumf}, CERN~\cite{cern}, 
BNL~\cite{aziz}, LBL~\cite{lbl}, and LAMPF~\cite{lanl}), at
energies common to those experiments (within $\Delta
E_\gamma$=5~MeV binning).  

%%%%%%%%%%%%%%%%%%%%%%%%%%%%%%%%%%%%%%%%%%%%%%%%%%%%%%%%%%%%%%
\section{FSI calculations}
\label{sec:FSI}

We extract the $\gnp\,$ \crs\ on free nucleon from the deuteron 
data in the quasi-free (QF) kinematical region of the $\gdpp\,$ 
reaction with fast knocked-out proton and slow proton-spectator 
assumed not to be involved in the pion production process. In 
this, so-called Impulse-Approximation (IA), the reaction 
mechanism corresponds to the diagram in Fig.~\ref{fig:g3}(a). 
There are 2 critical factors to be discussed when using this 
approach: 
\begin{enumerate}
	\item[1)] the neutron is bound and 
	\vspace{-2mm}
	\item[2)] there are $NN$- and $\pi N$-FSI effects.
\end{enumerate}

Item 1) means that effective mass of the neutron 
	$$m_{eff} = \sqrt{(p_d - p_s)^2} \approx m_n - \epsilon_d -
    \vec{p}_s^{~2}/m_N$$
is not equal to the mass of the free neutron $m_n$. Here,
$p_d$, $p_s$, $\vec{p}_s$, $\epsilon_d$, and $m_N$ are the deuteron
4-momentum, 4- and 3-momenta of the spectator, the deuteron binding
energy, and the nucleon mass, respectively.
% At $p_s\sim$150 MeV/$c$, in particular, the term $p_s^2/m_N\sim$25~MeV.
Simultaneously, the 
invariant mass $\sqrt{s_{\pi N}}$ of the final $\pi N$-system 
	$$\sqrt{s_{\pi N}} = \sqrt{s_{\gamma N}} 
   	= \sqrt{[(E_\gamma + m_d - E_s)^2 
   	- (\vec{p}_\gamma - \vec{p}_s)^2]}$$
depends on the proton-spectator momentum $\vec p_s$ ($s_{\gamma N}$ 
is the invariant mass squared of the initial $\gamma N$ state). Here, 
$E_{\gamma}$ ($E_s$), $m_d$, and $\vec p_{\gamma}$ are the total 
energy of the initial photon (proton-spectator), the deuteron mass, 
and the photon 3-momentum, respectively, and $E_{\gamma}=|\vec 
p_{\gamma}|$.

%virtuality of the neutron and on the direction of the
%proton-spectator ($s_{\gamma N}$ is the invariant mass squared of 
%the initial $\gamma N$ state). Here, $E_\gamma$ ($E_n$) is the 
%energy of the initial photon (virtual nucleon), $\vec {p}_\gamma$ 
%is the photon 3-momentum, $E_\gamma=|\vec {p}_\gamma|$.

%In our analysis, the $\gnp\,$ amplitude for a given $\sqrt{s_{\pi 
%N}}$ and centre-of-mass pion angle $\theta$ is assumed to be the 
%same as for the free neutron. At $p_s\sim$150 MeV/$c$, in particular,
%the mass shift $\approx \vec{p}_s^{ 2}/m_N\sim$25~MeV. The role of 
%the effect, perhaps, could be studied from comparison of the 
%$\gamma p\!\to\!\pi N$ amplitudes, obtained from $\gamma p$ and 
%$\gamma d$ experiments. 
Since $\sqrt{s_{\pi N}}$ depends on $\vec{p}_s$, the $\gamma 
N\!\to\!\pi N$ \crs\ extracted from the deuteron data with undetected 
nucleon-spectator is averaged over the energy range which depends on 
kinematical cuts for $\vec{p}_s$.   Thus, the ``effective" photon lab 
energy $E_{\gamma n}$, defined through the relation $s_{\gamma N}=
m^2_n+2 m_n E_{\gamma n}$ for the $\gamma n\to\pi^- p$ reaction, is 
smeared as well as the pion CM angle $\theta$ due the deuteron wave 
function. We estimated this smearing from simplified calculation, 
where $\gamma d\to\pi^- pp$ amplitude is proportional to the deuteron
wave function and depends only on the laboratory momentum of one of 
the final protons, say $p_2$, while $E_{\gamma n}$ is determined 
through the above-mentioned relation with the effective mass of the 
pion-proton pair with another proton $p_1$.

\begin{figure}[th]
\centerline{
\includegraphics[height=0.32\textwidth, angle=0]{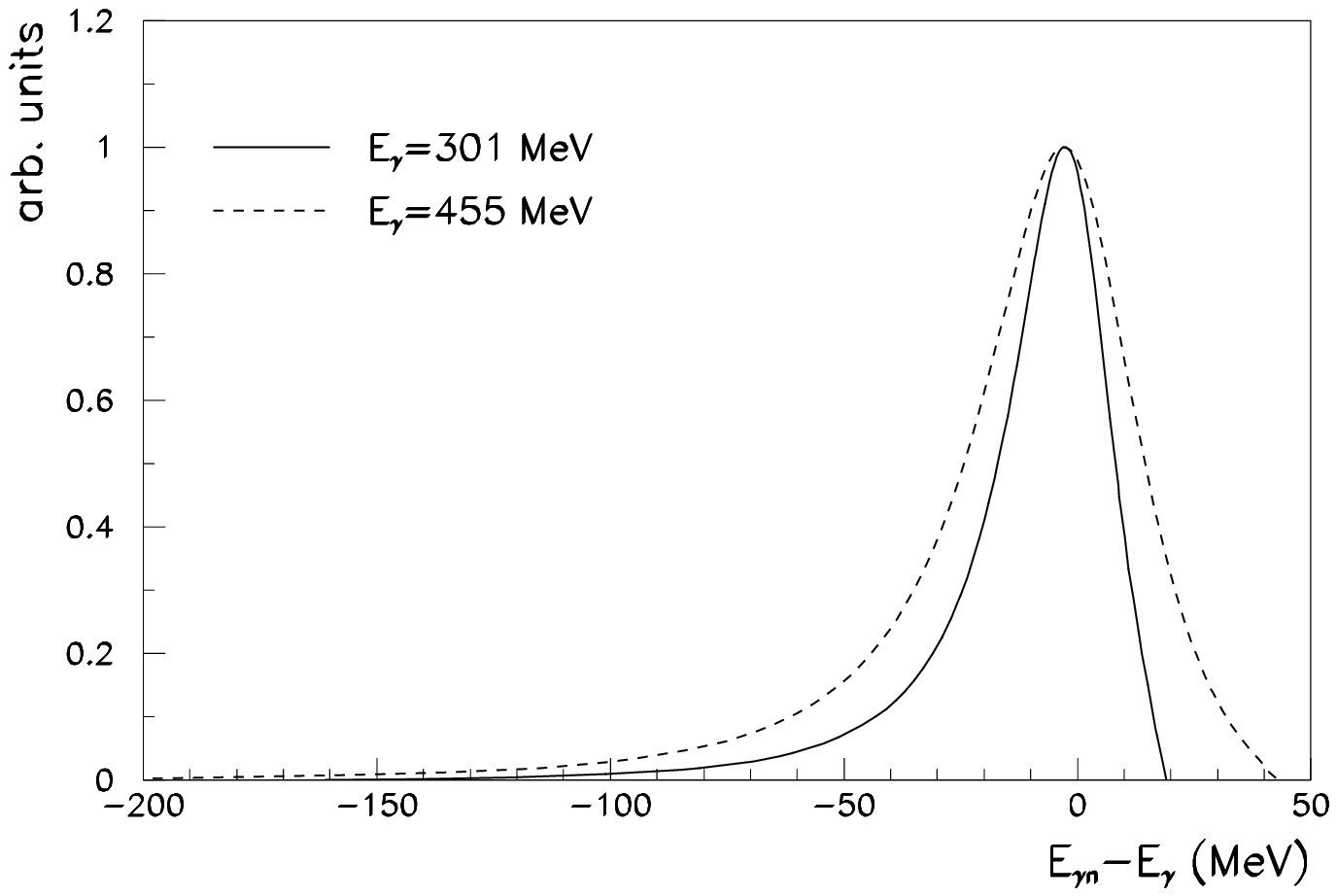}}
\caption{Distributions on the shift $\Delta E=E_{\gamma n}-E_{\gamma}$ 
	of the effective photon laboratory energy $E_{\gamma n}$ 
	on the neutron target at $E_{\gamma}=301$~MeV (solid 
	curve) and 455~MeV (dashed curve). The mean values are 
	$\sigma(\Delta)=-12\,(-14.6)$~MeV for $E_{\gamma}=301\,
	(455)$~MeV. \label{fig:g2}}
\end{figure}

\begin{figure}
\begin{center}
\includegraphics[height=2.3cm, keepaspectratio]{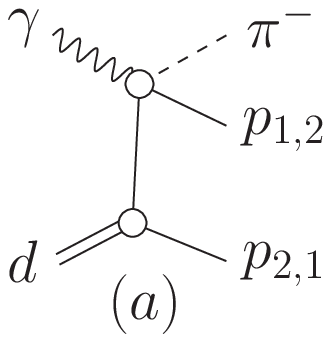}~
\includegraphics[height=2.3cm, keepaspectratio]{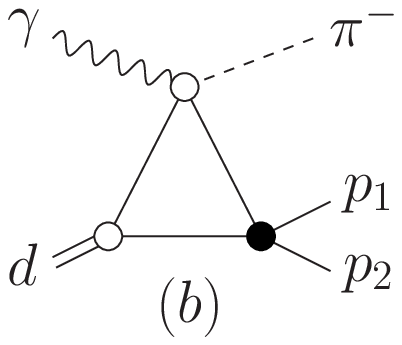}~~
\includegraphics[height=2.3cm, keepaspectratio]{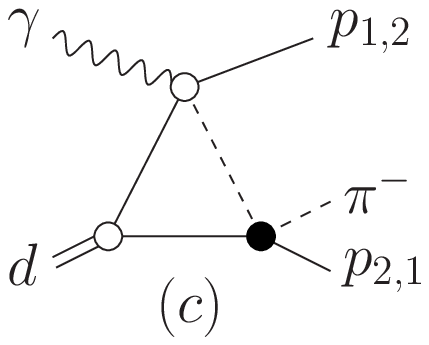}
\end{center}
\caption{Feynman diagrams for the leading terms of the $\gdpp$
        amplitude. (a) IA, (b) $pp$-FSI, and (c) $\pi$N-FSI.
        Filled black circles show FSI vertices.  Wavy, dashed,
        solid, and double lines correspond to the photons,
        pions, nucleons, and deuterons, respectively.}
        \label{fig:g3}
\end{figure}

%We estimated this smearing from simplified calculation 
%with $\gamma d\to\pi^- pp$ amplitude proportional to the deuteron 
%wave function.
%Fig.~\ref{fig:g2} shows the $E_{\gamma n}$ distributions for 
%$E_{\gamma}=301$ (curve 1) and 455~MeV (curve 2). The distributions 
%peak at $E_{\gamma n}$ close to $E_{\gamma}$ (by $\sim 3$~MeV to 
%the left). The mean values $\langle E_{\gamma n}\rangle$ and 
%dispersion $\sigma(E_{\gamma n})$ are $\langle E_{\gamma n}\rangle
%=289\,(440.4)$~MeV and $\sigma(E_{\gamma n})=20.3\,(32.2)$~MeV at 
%$E_{\gamma}=301\,(455)$~MeV. The dispersion $\sigma(E_{\gamma n})$ 
%exceeds $\sim\,$twice the 12-MeV intervals between the plots, i.e., 
%the neighbour plots are quite similar already due to smearing.
Fig.~\ref{fig:g2} shows distributions on $\Delta E=E_{\gamma n}-
E_{\gamma}$ at $E_{\gamma}=301$ and 455~MeV. The distributions peak 
at $\Delta\approx -3$~MeV, where $E_{\gamma n}$ is very close to
$E_{\gamma}$. The dispersion $\sigma(\Delta)=20.3\,(32.2)$~MeV at
$E_{\gamma}=301\,(455)$~MeV essentially exceeds the 12-MeV intervals 
between the plots on Fig.~\ref{fig:g1}. Thus, the neighbour plots on 
Fig.~\ref{fig:g1} are quite similar already due to smearing.
The plots on Fig.~\ref{fig:g1} also weakly depend on the energy 
$E_{\gamma}$ in the intervals $\sim\sigma(\Delta)$. Thus, the 
distortion of the extracted $\gamma n\to\pi^- p$ cross sections due 
to smearing effect is expected to be small.

%%%%%%%%%%%%%%%%%%%%%%%%%%%%%%%%%%%%%%%%%%%%%%%%%%%%%%%%%%
%%%%%%%%%%%%%%%%%%%%%%%%%%%%%%%%%%%%%%%%%%%%%%%%%%%%%
The results for the pion CM angle smearing in the angular range 
of MAMI-B data gives the mean values $\langle\theta\rangle$ very 
close to $\theta$, namely $|\langle\theta\rangle-\theta|<1^\circ$,
%$|\langle\theta\rangle-\sigma(\theta)|<1^o$, 
where the pion CM angles $\theta$ are obtained from $\theta_{lab}$ 
with the neutron at rest. The angle dispersion $\sigma(\theta)$
varies in the interval $\sim (3.0 - 5.5)^\circ$.

Item 2) corresponds to the inclusion of the FSI corrections.
Their leading terms correspond to Feynman diagrams shown on 
Fig.~\ref{fig:g3}(b,c).

\begin{figure*}[th]
\centerline{
\includegraphics[height=0.7\textwidth, angle=90]{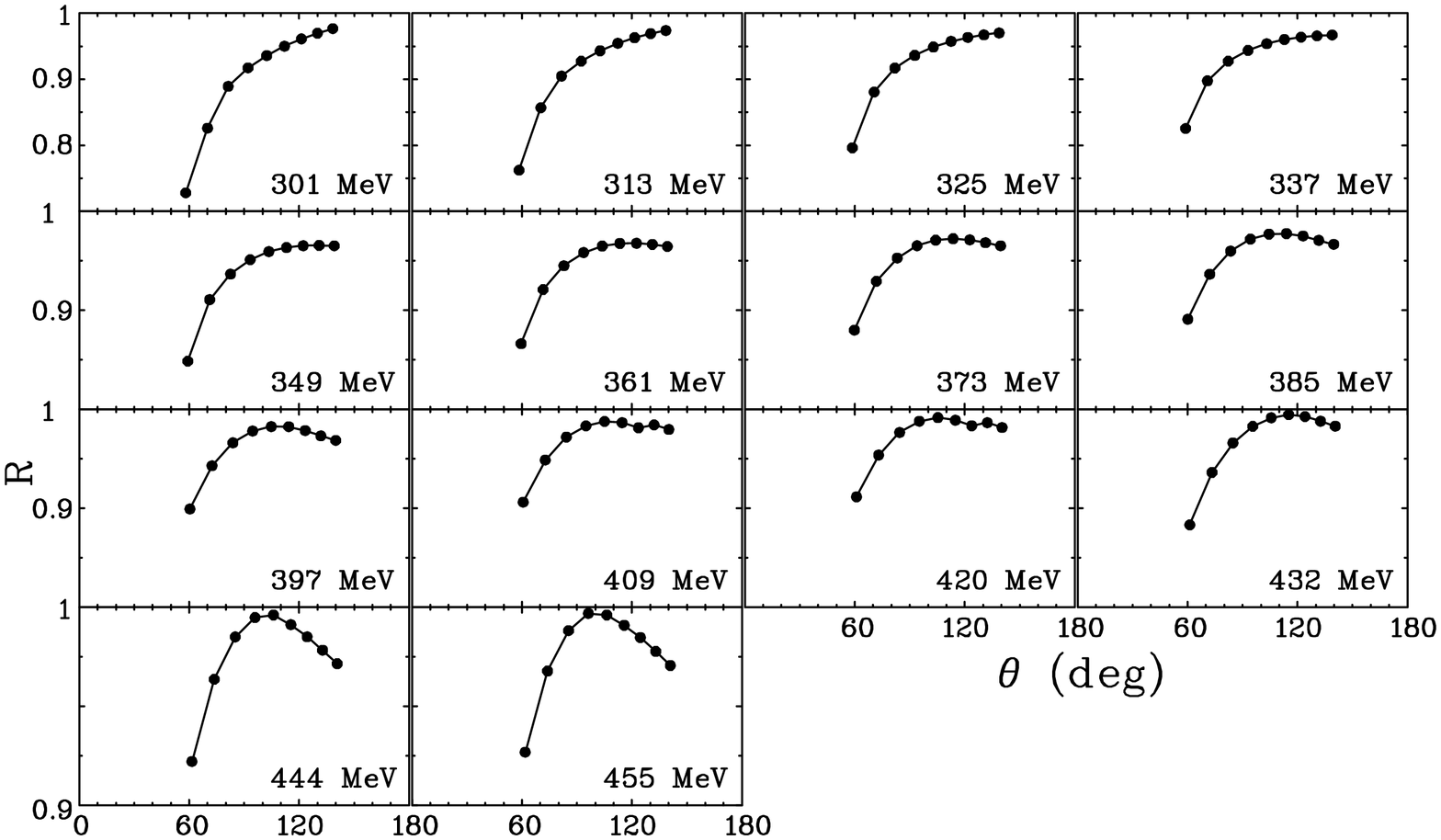}}
\caption{FSI correction factor $R$ for $\gnp$ as a function
        of $\theta$, where $\theta$ is the production angle of
        $\pi^-$ in the CM frame.  The present calculations
        (solid circles) are shown for 14 energy bins.
	There are no uncertaintes given. Curves may help
	to lead eyes. \label{fig:g4}}
\end{figure*}

Calculations of the $\gdpp\,$ differential cross sections, with 
the FSI taken into account (all the diagrams on Fig.~\ref{fig:g3}, 
were included) were done as we did recently~\cite{FSI,data} for the 
CLAS data ($E_\gamma = 1050 - 2700~MeV$ and $\theta = 32 - 
157^\circ$)~\cite{CLAS}.  The SAID phenomenological amplitudes for  
$\gamma N\!\to\!\pi N$~\cite{pr_PWA}, $NN$-elastic~\cite{NN_PWA}, 
and $\pi N$-elastic~\cite{piN_PWA} were used as inputs to calculate 
the diagrams in Fig.~\ref{fig:g3}.  The Bonn potential~\cite{Bonn} 
was used for the deuteron description. In Ref.~\cite{data}, we 
calculated the FSI correction factor $R$ dependent on E$_\gamma$ 
and $\theta$ (see details in Refs.~\cite{FSI,data}), and fitted 
these CLAS data vs. the world $\gamma N\to\pi N$ database~\cite{SAID} 
to get new multipoles and determine resonance EM couplings~\cite{data}. 
The FSI corrections for the CLAS "quasi-free" kinematics were found 
to be small, as mentioned above. Our FSI calculations were done
\cite{FSI} over the broad energy range (threshold to 2700~MeV for 
$E_\gamma$) and for the full angular coverage ($\theta = 0 - 
180^\circ$). As an illustration, Fig.~\ref{fig:g4} shows the FSI 
correction factor $R=R(E_\gamma,\theta)$ for the present $\gnp$ 
differential \crss\ as a function of the pion production angle in 
the CM($\pi^-p$) frame, $\theta$, for different energies over the 
range of the MAMI-B experiment.  Overall, the FSI correction factor 
$R< 1$, while the effect, i.e., the $(1-R)$ value varied from 10\%
to 30\% depending on the kinematics and the behavior is very smooth 
vs. pion production angle.  Note that $R(E_{\gamma},\theta)$ is the 
FSI correction factor for the $\gamma n\to\pi^- p$ cross section 
averaged over the lab photon energy $E_{\gamma n}$. Fig.~\ref{fig:g3} 
shows that $R$ depends slowly on the energy in the intervals 
$\sim\sigma(\Delta)$. Thus, smearing effect illustrated on 
Fig.~\ref{fig:g2}, weakly affects the FSI correction procedure for
the extracted $\gamma n\to\pi^- p$ cross section at a given energies.

The contribution of FSI calculations~\cite{FSI} to the overall
systematics is estimated to be 2\%.

%%%%%%%%%%%%%%%%%%%%%%%%%%%%%%%%%%%%%%%%%%%%%%%%%%%%
%%%%%%%%%%%%%%%%%%%%%%%%%%%%%%%%%%%%%%%%%%%%%%%%%%%%%%%%

%%%%%%%%%%%%%%%%%%%%%%%%%%%%%%%%%%%%%%%%%%%%%%%%%%%%

%%%%%%%%%%%%%%%%%%%%%%%%%%%%%%%%%%%%%%%%%%%%%%%%%%%%%
\section{Results}
\label{sec:res}

In fitting the database, $\chi^2$ is calculated using
\be
	\chi^2 = \sum^{N_{data}}_{i} \left( {{O_i - N_j O_i^{exp}}\over
	{\delta O_i}} \right)^2 + \sum_{j}^{N_{dist}} \left( {{N_j - 1 }\over
	{\delta N_j }} \right)^2,
\label{chi}
\ee
where $O_i$ and $O_i^{exp}$ are calculated and experimental 
observables, for a given energy and angle, and $\delta O_i$ is the 
statistical uncertainty. The systematic error, $\delta N_j$, for a 
given angular distribution, is used to calculate a second 
contribution to $\chi^2$ due to overall normalization ($N_j$) of 
angular datasets.

In Fig.~\ref{fig:g1}, which compares the present measurements with
corresponding results derived from pion-induced reactions, no FSI 
(nor any data renormalization) corrections have been applied. The 
curves are predictions from SAID and MAID and are generally quite 
consistent. Without corrections, the pion- and photo-induced data 
are reasonably consistent where they can be compared (the 
pion-induced results having significantly larger uncertainties). 
Comparisons with the SAID and MAID predictions show reasonable 
agreement in terms of shape but, at a number of energies, there is 
a clear difference in the overall normalization. 

\begin{figure*}[th]
\centerline{
\includegraphics[height=1\textwidth, angle=90]{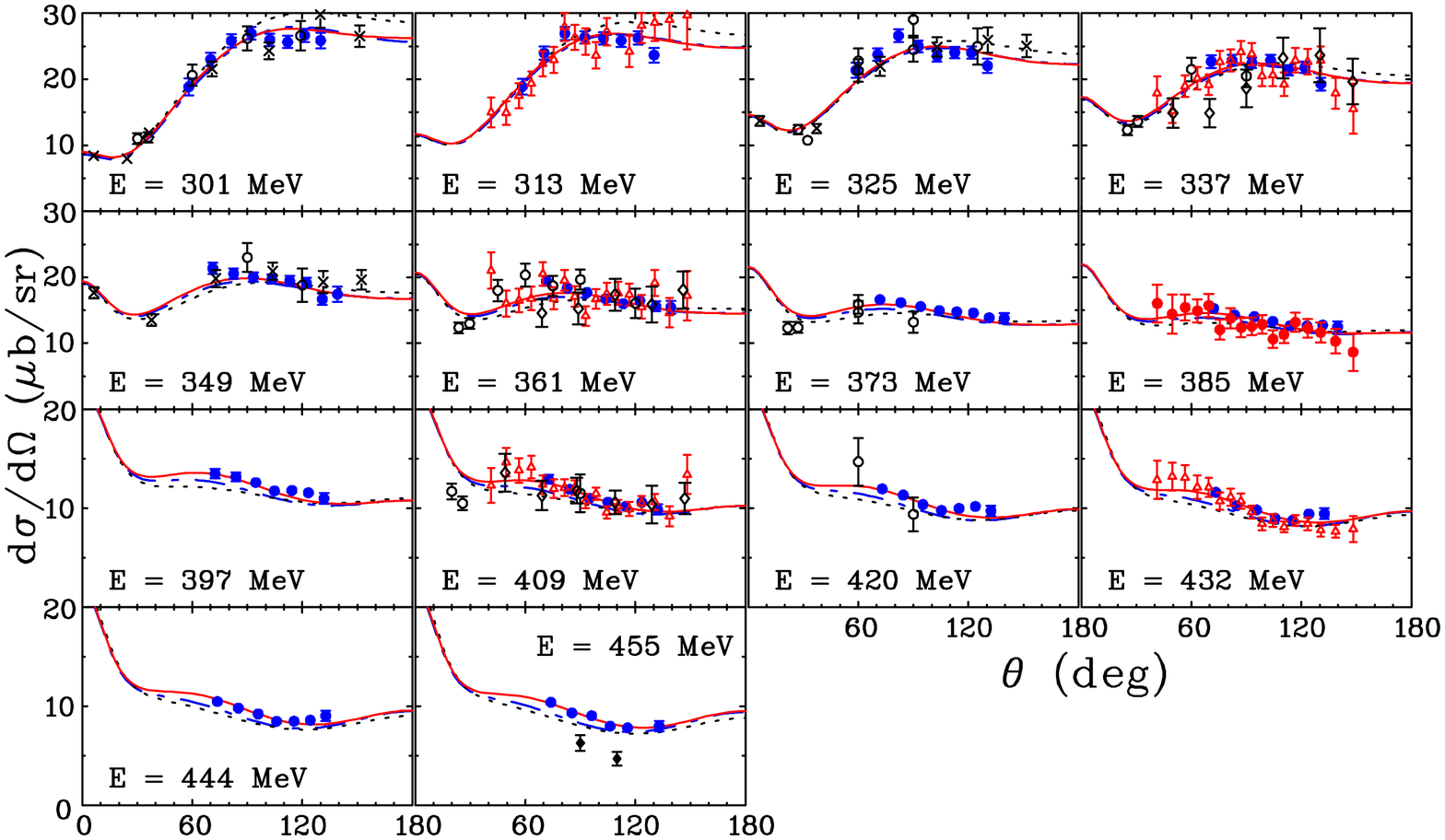}}
\caption{(Color online) Differential \crss\ for $\gnp$ as a
        function of $\theta$, where $\theta$ is the production
        angle of $\pi^-$ in the CM frame. Notation of data and 
	solutions are the same as in Fig.~\protect\ref{fig:g1}.
	MAMI-B data including in the PE12 fit (red solid line). 
	\label{fig:g5}}
\end{figure*}

In Fig.~\ref{fig:g5}, both FSI and data renormalization have been 
applied. A solid curve, giving the result of a fit, is compared 
with the aforementioned predictions. The data renormalization, 
required for a best-fit result, changes from an average of 7\%, 
for the SAID prediction, to about 4.5\%,after the data have been 
included in the fit. In both cases, the second term in Eq.~(\ref{chi}) 
above (due to renormalization) contributes nearly 50\% to the total 
chi-squared.  

Changes to the multipoles in the revised fit are small. The dominant 
multipole contribution from the $\Delta (1232)$ is not changed 
significantly, as one would expect. Together with the measurements 
of Ref.~\cite{data}, we now have a nearly complete coverage of the 
resonance region for unpolarized cross sections, in this reaction. 
Further progress will require polarized measurements, which are 
expected from the CLAS Collaboration~\cite{dar}.

Since the MAMI-B results for the $\gnp$ differential \crss\ 
consist of $126$ experimental points [$E_\gamma = 301 - 
455~MeV$ and $\theta = 58 - 141^\circ$], they are not 
tabulated in this or the previous~\cite{MAMI} publication, 
but are available in the SAID database~\cite{SAID} along 
with their uncertainties and the energy binning.  

The $\chi^2$ contribution of MAMI-B data (including FSI corrections) 
is $\chi^2$/data = 249.4/104 = 2.4 while, prior to fitting, for 
SN11~\cite{sn11} (MAID07~\cite{MAID}), we had $\chi^2$/data = 
605.8/104 = 5.8 (623.2/104 = 6.0).

The MAMI-B data (including FSI corrections) and the results 
from hadronic data appear to agree well at these energies 
(Fig.~\ref{fig:g4}).  In particular, the $\chi^2$ contribution 
from recent CB$@$BNL~\cite{aziz} and MAMI-B measurements at 6 
overlapped energies [313, 337, 361, 385, 409, and 432~MeV] 
results $\chi^2$/data = 97.7/102 = 1.0 and $\chi^2$/data = 
103.9/45 = 2.3, respectively, for the PE12 solution.  While for 
previous SN11 solution~\cite{sn11}, it gives $\chi^2$/data = 
87.4/102 = 0.9 and $\chi^2$/data = 188.6/45 = 4.2, respectively 
(MAMI-B data had no FSI corrections).
%%%%%%%%%%%%%%%%%%%%%%%%%%%%%%%%%%%%%%%%%%%%%%%%%%%%%

%%%%%%%%%%%%%%%%%%%%%%%%%%%%%%%%%%%%%%%%%%%%%%%%%%%%%%%%%%%%%%%%%%%
\section{Summary and Conclusion}
\label{sec:conc}

A comprehensive set of differential \crss\ at 14 energies for 
negative-pion photoproduction on the neutron, via the reaction 
$\gdpp$, have been determined with a MAMI-B tagged-photon beam 
for incident photon energies from 301 to 455~MeV. To accomplish 
a state-of-the-art analysis, we included FSI corrections 
using a diagrammatic technique, taking into account a 
kinematical cut with momenta less (more) than $\sim$270~MeV/$c$ 
for slow (fast) outgoing protons.

On the experimental side, further improvements in the PWAs 
await more data, specifically in the region above 1~GeV, where 
the number of measurements for this reaction is small.  Of 
particular importance in all energy regions is the need for 
data obtained involving polarized photons and/or polarized targets.  
Some of these data are already available in Ref.~\cite{data}.
Due to the closing of hadron facilities, new $\pi^-p\!\to\!\gamma 
n$ experiments are not planned and only $\gnp$ measurements 
are possible at electromagnetic facilities using deuterium 
targets. Our agreement with existing $\pi^-$ photoproduction 
measurements leads us to believe that these photoproduction 
measurements are reliable despite the necessity of using a 
deuterium target. 

Obviously, any meson photoproduction treatment on the 
``neutron" target requires a FSI study. Generally, FSI depends 
on the full set of kinematical variables of the reaction. In 
our analysis, the FSI correction factor depends on the photon 
energy, meson production angle, and is averaged on the rest of 
variables in the region of ``quasi-free" process on the neutron.

%%%%%%%%%%%%%%%%%%%%%%%%%%%%%%%%%%%%%%%%%%%%%%%%%%%%%%%%%%%%%%%%%
\vspace{0.5in}
\begin{acknowledgments}
We acknowledge the outstanding efforts of the GDH and A2 Collaborations 
who made the experiment possible. This work was supported in 
part by the U.S. Department of Energy Grant No.~DE--FG02--99ER41110, 
by the Russian RFBR Grant No.~02--0216465, by the Russian Atomic 
Energy Corporation ``Rosatom" Grant No.~NSb--4172.2010.2, and 
the Italian Istituto Nazionale di Fisica Nucleare.
\end{acknowledgments}

%%%%%%%%%%%%%%%%%%%%%%%%%%%%%%%%%%%%%%%%%%%%%%%%%%%%%%%%%%%%%%%%%%%%%%

\end{document}